# Experimental study of a transformer with superconducting elements for fault current limitation and energy redistribution

V. Meerovich[*], V. Sokolovsky,

Physics Department, Ben-Gurion University of the Negev, P.O. Box 653, Beer-Sheva, Israel

*Abstract*

Numerous proposed and developed superconducting fault current limiters and self-limiting transformers limit successfully fault currents but do not provide uninterrupted supplying of consumers. A design investigated in the work combines the functions of a conventional transformer with the functions of fast energy redistribution and fault protection. The device constitutes a transformer containing an additional high-temperature superconducting (HTS) coil short-circuited by a thin film HTS switching element. Fault current limitation and redistribution of the power flow to a standby line are achieved as a result of a fast transition of the superconducting switching element from the superconducting into the normal state. Transient and steady-state characteristics were experimentally investigated. A mathematical model of the device operation was proposed, and the calculated results were found to be in good agreement with the experimental data. The application field and basic requirements to such devices were discussed and it was shown that the proposed device meets these requirements.

## 1. Introduction

In order to increase the reliability of the electric power consumption and to protect electric equipment during fault events, the devices are needed that can be activated in less than a half-period of AC current, limit fault current and redistribute power to a standby line. Many various designs of fault current limiters (FCL) were proposed, among them the devices based on the properties of the superconducting-normal state transition (S-N transition). Inductive, resistive and various "hybrid" designs of superconducting FCLs were considered in different modifications, based on different superconducting materials and well described in numerous reviews and reported at international conferences [1-5]. Several high power prototypes of superconducting FCLs have been built and successfully tested showing feasibility of various proposed concepts for application in power electric systems [6-11]. Besides FCLs, a fault current limiting transformer [12] and a transformer operating as a circuit-breaker under a fault [13] were proposed. However, all these devices do not prevent breaks in power supply of consumers during faults. A power interruption even during very short faults (0.1-0.2 s) is inadmissible for many long-term technological processes.

---

[*] Corresponding author. Tel.: +972-8-647-2458; fax: +972-8-647-2903.
*E-mail address*: victorm@bgu.ac.il (V. Meerovich).

One conventional method for protection against unexpected breaks in power supplying is using two or more power lines leading from different substations and connected to a load through controlled AC-DC or AC-AC converters. During a fault in one line, this line is disconnected by the converter electronics and further power service is realized through undamaged lines. An alternative to building of additional power lines is the application of energy storage systems using analogous converters for coupling to a power network [14, 15].

Another approach to the problem of faults and consumption breaks is the development of multi-functional fast operating systems combining current-limitation and energy distribution functions, providing a fast switching of power flow to a standby line during faults.

This paper is aimed to present the results of the study of the operation of a novel device combining the functions of a conventional transformer with the functions of fast power redistribution and fault protection.

## 2. Design and operation principle

We studied the operation of a device shown schematically in Fig. 1, which constitutes a transformer containing additional coils, one of them is a high-temperature superconducting (HTS) coil short-circuited by a switching HTS element. The operation of the device is based on a fast S-N transition in the switching element.

The transformer consists of a primary coil $W_1$, a two-sectional secondary coil with sections $W_2$ and $W_2$', a standby secondary coil $W_3$ and a superconducting coil $W_4$ placed on the magnetic leg with coils $W_2$' and $W_3$. The sections $W_2$ and $W_2$' are located on different magnetic legs and counter-connected in series so that the voltages induced in the coils are of opposite signs. The superconducting coil is short-circuited by an HTS switching element.

Under normal operating conditions, the switching element is in the superconducting state and the coil $W_4$ compensates the magnetic flux in the magnetic branch with the coils $W_2$' and $W_3$. As a result of it, any electromotive forces is not induced in the coils $W_2$' and $W_3$. The device operates as a usual transformer with the primary coil $W_1$ and the secondary coil $W_2$ supplying a consumer. When an excess current caused by an overload or a fault appears in line 1, the current in the section $W_2$' increases resulting in the current increase in the coil $W_4$ and in the switching element. This initiates the transition of the switching element to the normal state. Electromotive forces appear in the coils $W_2$' and $W_3$. The numbers of turns in the coils $W_2$ and $W_2$' are chosen so that their induced electromotive forces are compensated. The voltage of the secondary coil $W_2$-$W_2$' decreases drastically and the fault current in line 1 is limited. At the same time, standby line 2 is switched on. Thus, the device performs simultaneously two functions: limitation of the current in the line where the fault occurs, and energy redistribution providing the uninterrupted energy supply to a consumer.

The considered design differs from a transformer-circuit-breaker proposed by Y. Bashkirov et al. [13] by introducing an additional secondary coil. It allows one to expand the functions of the device: along with the fault current limitation, the device provides the continuity of power service due to redistribution of power flow to a standby line.

## 3. Experimental model

The experimental model of the device was built with the following parameters: the turn numbers of in the coils: $W_1$ - 100, $W_2$ – 50, $W_2$' - 88, $W_3$ - 100. The HTS coil $W_4$ had 8 turns and was short-circuited by a HTS switching element. The coils were placed on an E-type core with dimensions 150x165x30 mm$^3$.

A Bi-2223/Ag multifilamentary tape used for manufacturing of the HTS coil was fabricated by the oxide powder-in-tube method [16]. The tape was wound on a cylindrical former with the external diameter of 52 mm and the height of 38 mm. The measurements by the 4-point probe method gave the critical current of 18 A determined with 1 µV/cm criterion. Under AC conditions, the active component of the voltage drop across the coil was equal to 20 µV/cm at the current amplitude of 22 A and at the temperature of 77 K.

The switching element was fabricated on the base of HTS YBCO epitaxial thin film of 300 nm in thickness (protected by a gold layer, 100 nm in thickness) deposited by thermal reactive co-evaporation on a sapphire substrate of 0.5 mm in thickness [11]. The switch was fabricated in the form of a strip with two broad current terminals, active length of 18 mm and width of 2.5 mm. The critical temperature of the film was about 87 K, the critical current of the switch - 10 A at 77 K.

The thin film switch and the HTS coil were connected using silver contacts. Resistance of the contacts was less than 0.006 Ω. The operation of the HTS coil with the thin film switching element was previously investigated in a model of an inductive FCL [17].

## 4. Experimental results

The testing of the experimental model was carried out in two stages. The secondary coils were connected to different resistors which simulated loads in the main and standby lines. In the first stage, the waves of voltages and currents were recorded in two operating steady-state modes: under normal conditions (Fig. 2) and a fault in line 1 (Fig. 3). Under normal conditions, the switching element was in the superconducting state (at a current below the critical value). The voltage induced in the coil $W_3$ is low and appears only due to imperfect compensation of the magnetic flux by the superconducting coil. A fault in line 1 was modeled by short-circuiting of the load in the line. As a result of it, the current in the section $W_2$' increased, the switch *was transferred* into the normal state and the voltage across the secondary coil $W_2$-$W_2$' was decreased and determined by the impedance of the short-circuited line (Fig. 3a). The limited fault current was about twice the nominal current (Fig. 3c). The fault current in line 1 without limitation would be about 20 times more than the nominal current. As one can see from the comparison of Figs. 2a and 3a, the voltage supplying standby line 2 during a fault achieved almost the nominal value that was under normal conditions in line 1.

Note that in Fig. 2 the voltages across the primary and secondary coils are about in opposite phases while the currents have a shift. It is explained by a relatively low magnetizing reactance in our experimental model and relatively high load resistance and leakage reactance. In full-scale devices, the magnetizing reactance is very high, and the currents must be also in opposite phases.

The second stage of the testing was related to investigation of the dynamic characteristics of the device. A short-circuit regime in line 1 was performed by an electronic switch connected in parallel to the load. A variable resistor was inserted in the circuit in series with the switch. Changing the resistance, different cases of remoteness of a fault point from the transformer were modeled.

Fig. 4 presents typical scope traces of transients in the case of remote fault *(the fault current was the least)*. Fault current limitation as well as the appearance of the voltage across the coil $W_3$ supplying standby line 2 occurred without any delay (Fig. 4c). The recovery of the normal operation of the transformer after a fault clearing was observed in less than a half-period.

Note that, during a fault, the device effectively protects the supplying line connected to the primary coil: the voltage and current in the primary coil remain practically the same (Fig. 4a).

## 5. Mathematical model of the device

The mathematical model of the transformer device as a power system element is based on the equations of magnetic coupled circuits:

$$u_1 = R_{11}i_1 + L_{11}\frac{di_1}{dt} + (L_{21} - L_{2'1})\frac{di_2}{dt} + L_{31}\frac{di_3}{dt} + L_{41}\frac{di_4}{dt}$$

$$0 = (L_{12} - L_{12'})\frac{di_1}{dt} + R_{22}i_2 + (L_{22} - 2L_{2'2} + L_{2'2'})\frac{di_2}{dt} + R_2 i_2 + L_2\frac{di_2}{dt}$$

$$+ (L_{32} - L_{32'})\frac{di_3}{dt} + (L_{42} - L_{42'})\frac{di_4}{dt} \tag{1}$$

$$0 = L_{13}\frac{di_1}{dt} + (L_{23} - L_{2'3})\frac{di_2}{dt} + R_{33}i_3 + L_{33}\frac{di_3}{dt} + R_3 i_3 + L_3\frac{di_3}{dt} + L_{43}\frac{di_4}{dt}$$

$$0 = L_{14}\frac{di_1}{dt} + (L_{24} - L_{2'4})\frac{di_2}{dt} + L_{34}\frac{di_3}{dt} + L_{44}\frac{di_4}{dt} + u_s$$

where $u_1$ is the primary terminal voltage; $i_i$, $R_{ii}$ and $L_{ii}$ are the current, resistance and inductance of the coil $W_i$ ($i = 1, 2, 2`, 3, 4$); $R_i$ and $L_i$ is the resistance and inductance of the load connected to the secondary coil $W_i$ ($i = 2, 3$); $L_{ij}$ is the mutual inductance of the coils $W_i$ and $W_j$. Because the sections $W_2$ and $W_2$' are electrically connected, we put $i_2 = i_{2'}$. The voltage drop $u_s$ across the superconducting switching element depends on the current and temperature. A frequently used approximation for the voltage-current characteristic of a superconductor [5, 18] is:

$$u_s = \begin{cases} 0 & \text{if } |i_s| \le i_c(T) \text{ and } T < T_c \\ R_\rho[i_s - \text{sign}(i_s)i_c(T)] & \text{if } |i_s| > i_c(T) \text{ and } T < T_c \\ R_n i_s & T \ge T_c \end{cases} \tag{2}$$

where $R_\rho$ is the resistance of the superconductor in the resistive state (so-called the flux flow resistance), $R_n$ is the resistance of the superconductor in the normal state, $i_s$ is the instantaneous value of the current in the superconductor, and $i_c$ is its critical current which depends on the temperature $T$, $T_c$ is the critical temperature of the superconductor.

The transient process is described by the simultaneous solution of Eqs. (1), (2) and the heat equation for the superconducting element. The character of the transient process after $i_s$ achieves the critical value is determined by the relationship between the rate of heating of the superconducting element and the rate of the transient current attenuation.

At quick heating of the superconductor, the condition required for a full-scale device, processes of the fault and S-N transition can be considered as a stepwise change of the load in the main line ($R_2$ and $L_2$ go down to zero) and of the resistance of the switching element (from zero to $R_n$). A peculiarity of the considered device is that transient currents attenuate quickly due to relatively large resistances $R_n$ and $R_3$. This is well illustrated by our experiments with the device model: scope traces in Fig. 4 show that the transient processes are very short and the device operation can be simulated as a jump from one steady state to another. Based on this conclusion, two operating regimes were simulated: 1) normal operation when the switching element is in the superconducting state ($R_{44} \sim 0$); 2) fault when $R_2$ is low and the switching element is in the normal state with a high resistance. The simulation results for the parameters of our experimental model and for the amplitude of the primary voltage $U_1$ = 15V are shown in Figs. 2b and 3b for normal and fault conditions, respectively. These results are in good agreement with the experimental data presented in Figs. 2a and 3a.

## 6. Discussion

The obtained experimental results have shown the feasibility of the proposed design: the limitation effect and power redistribution were achieved without any delay, in time less than ¼ of AC period. The limitation was achieved without overvoltages in the circuit. The S-N transition occurred only in the HTS switch while the secondary HTS coil remained in the superconducting state in all the regimes. Non-sinusoidal experimental traces of the voltage and current (Fig. 3) indicate that the resistance of the switch was changed markedly during every AC period due to heat processes. The primary current $I_1$ did not practically change due to fault event while the secondary current $I_2$ increased more than twice during fault conditions (Fig. 4). Emphasize, that this increase is required for the successful operation of the protecting automatics, which has to "see" a fault and to send the command to the circuit breaker to open the circuit. Note also, that compensation of the electromotive forces induced in two sections $W_2$ and $W_2$' under fault conditions cannot be perfect, even if amplitudes of $U_2$ and $U_2$' are equal. This is explained by asymmetry of the device resulting to the phase shift between the magnetic fluxes in the central and side legs of the magnetic core. Therefore the electromotive forces induced in the sections $W_2$ and $W_2$' are also shifted. Figs. 4a and 4b demonstrate the recovery of the device to the initial conditions in 3-4 periods after a fault clearing.

The full-scale device should be built using multifilamentary HTS wires to decrease AC losses in the superconducting coil. The switching element can be fabricated as a set of HTS films connected in parallel and in series similarly to switching elements of FCLs [5, 8, 17]. An important technological problem is to lower the contact resistance between a HTS wire and films. There are three conditions determining the accessible value of this resistance. First, the contacts have not to be sources of normal zone nucleation. Second, the total resistance of the contacts has to be much less than the impedance of the secondary HTS coil. Third, the restriction on the resistance value is dictated by the economically sound level of power losses under a normal regime of the protected circuit.

If the current in a line is higher than double the maximum of the rated current in the normal regime, this regime is considered as a fault. This condition determines the required critical current of the switching element: when the instantaneous current in the circuit of the coils $W_2$ and $W_2$' exceeds twice the nominal current amplitude, the current in the superconducting switch has to achieve the critical value.

The critical current can be calculated using equations (1). However, with satisfactory accuracy, it is determined by the ratio of the turn numbers in the coil $W_{2'}$ and the superconducting coil.

The resistance $R_n$ of the switching element in the limitation regime is determined by several conditions. The first condition requires that the superconducting coil $W_4$ does not influence on the magnetic flux distribution when the superconducting switch is in the normal state. It is achieved if:

$$g = R_n / X_{44} \gg 1. \qquad (3)$$

The estimations show that $g = 5$ is sufficient [5,17]. Under condition (3), the reduced resistance of the switch in the normal state is much higher than all the load impedances. Therefore, almost whole of the power goes to the load through the standby line.

The second limiting condition for $R_n$ follows from the restrictions on dissipated energy to prevent overheating of the switching element. The dissipated energy during the time of a fault, $t_k$, is mainly determined by the turns ratio of the primary and superconducting coils:

$$\left[ \frac{U_1 W_4}{2\sqrt{2} W_1} \right]^2 \frac{t_k}{R_n} \leq Q_{max} , \qquad (4)$$

where $Q_{max}$ is the maximum admissible dissipated energy in the switching element.

Dimensions of the switching element can be evaluated using the formula given in [5, 17] for fault current limiters. It is expected that the parameters of superconducting elements will be close to the parameters of these elements for an inductive FCL of the same power. For example, the parameters of 11kV/600A superconducting FCL were presented in [5, 17]. To calculate the parameters of the electromagnetic system, the methods developed for conventional transformers can be apply [19, 20]. It should be noted that, for our transformer device, there is no a short-circuit regime, such as for conventional transformers: the primary current and voltage do not practically change at the fault in the main line (Fig. 4). Thus, the device protects the primary circuits of transformers against the current increase and voltage reduction during faults: the property unavailable for other fault current limiters.

The proposed transformer device can be used for uninterrupted supply of long-term technological processes. Economical gain from the device application should be determined from the comparison with other solutions as building an additional substation or using of energy storage systems. The proposed device uses the same converters which are needed for conventional methods. The main and standby lines are connected to the input terminal of a converter which disconnects the fault line. An additional advantage of the proposed device, in the comparison with the conventional methods, is the limitation of fault currents in both the primary circuit and the faulted line.

The results of our investigations have shown that the proposed device meets the basic requirements following from the above considered application: quick (in less than ¼ of AC period) switching of the power between the main and standby lines and quick (a few periods) recovery to the initial conditions after a fault clearing. The switching between different modes occurs without dangerous overvoltages.

## 6. Conclusion

The performed investigations have confirmed the feasibility of the proposed design. The experiments have shown that the system "superconducting coil – switch" can be used as a module for controlling the magnetic flux distribution in transformer devices. It opens ways for developing designs realizing the fault current limitation simultaneously with the redistribution of power flows to standby lines. This feature provides uninterrupted supply of consumers even during a fault event.

*Acknowledgements* - This research was supported by the Ministry of Infrastructure of Israel through the Grant "Development of Transformer Device with Superconducting Coil for Controlling Current and Energy Distribution in Electric Power Network", 2001-2004. The high temperature superconducting thin films were supplied from Siemens AG, Corporate Technology Erlangen, Germany. The authors wish to thank Dr. W. Schmidt and B. Utz for fabrication of the experimental samples.

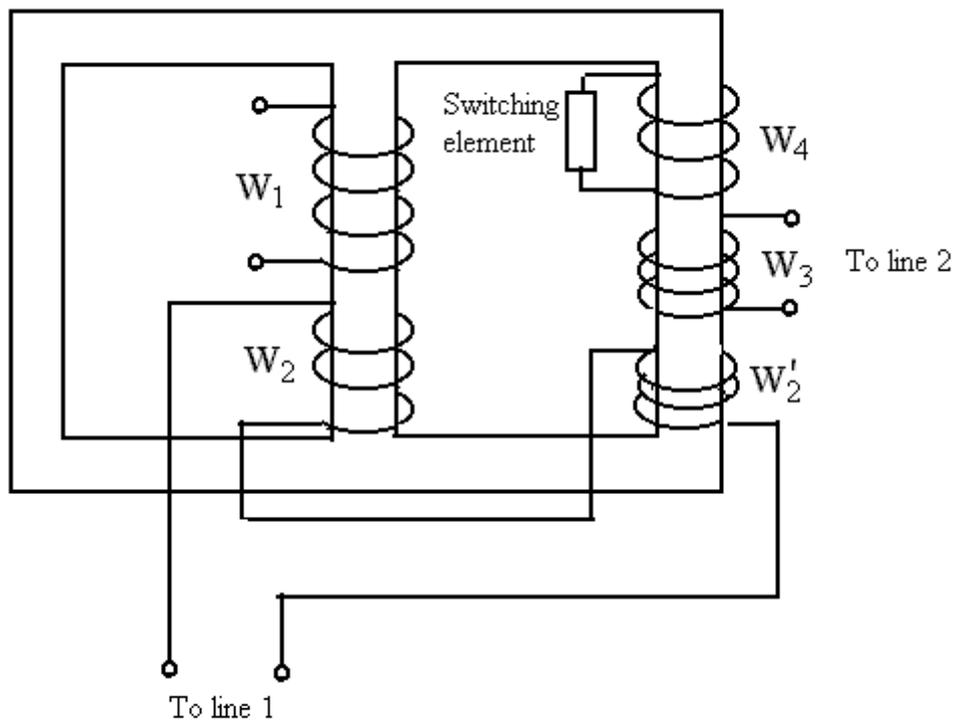

Fig.1. Schematic view of the experimental model.

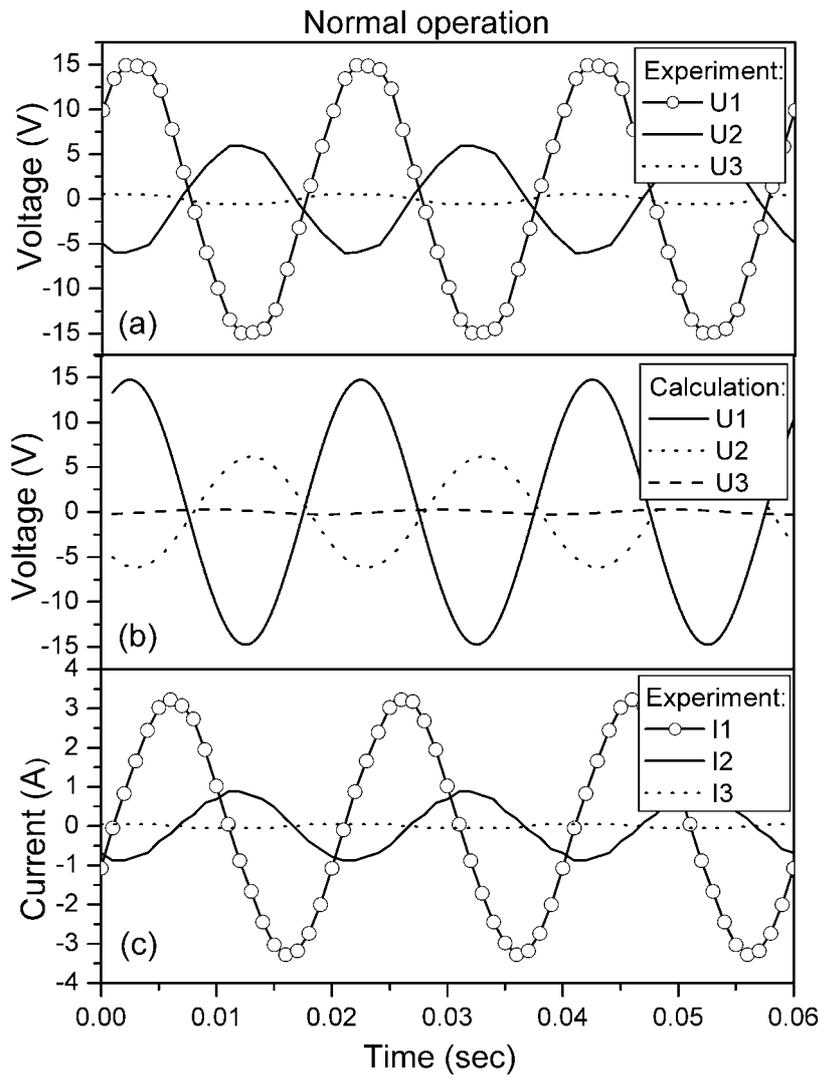

Fig. 2. Normal operating conditions (HTS switch is in the superconducting state): (a) – experimental waveforms of voltages; (b) – calculated waveforms of voltages at $R_2 = R_3 = 10$ Ohm, $R_4 = 0.006$ Ohm; (c) experimental waveforms of currents. $U_1$, $I_1$ - voltage and current in the primary coil; $U_2$, $I_2$ – the main secondary coil; $U_3$, $I_3$ – the standby secondary coil.

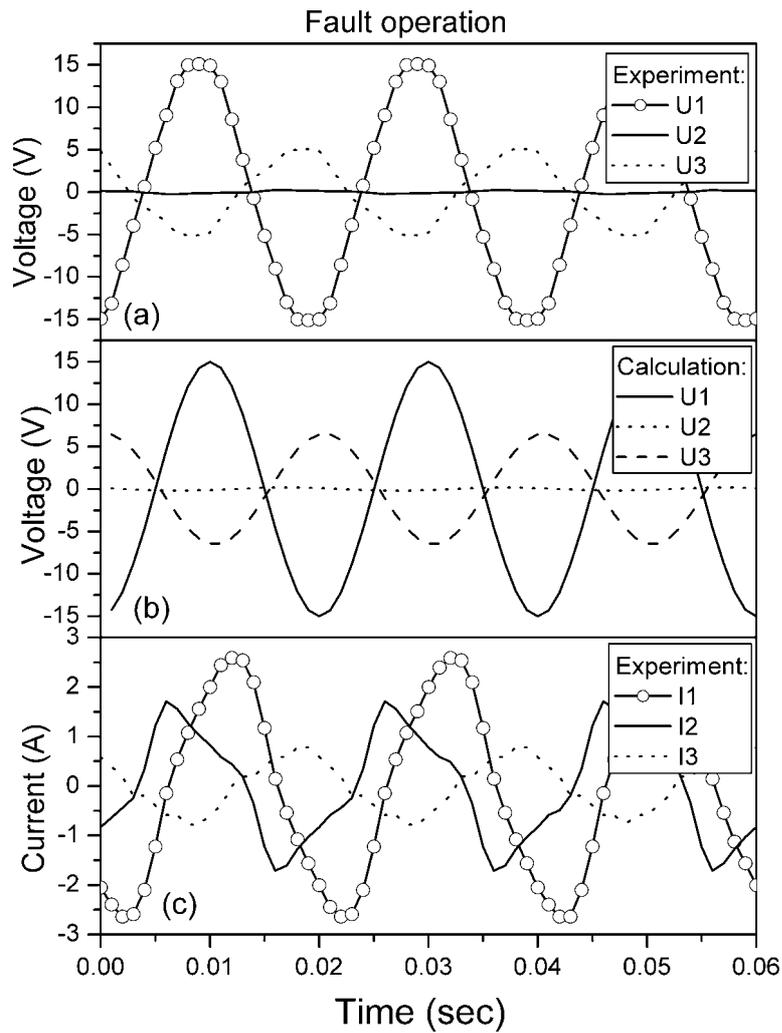

Fig. 3. Fault (HTS switch is in the normal state above the critical temperature of S-N transition): (a) – experimental waveforms of voltages; (b) – calculated waveforms of voltages at $R_3$ = 10 Ohm, $R_4$ = 100 Ohm; (c) experimental waveforms of currents. $U_1$, $I_1$ - voltage and current in the primary coil; $U_2$, $I_2$ – the main secondary coil; $U_3$, $I_3$ – the standby secondary coil.

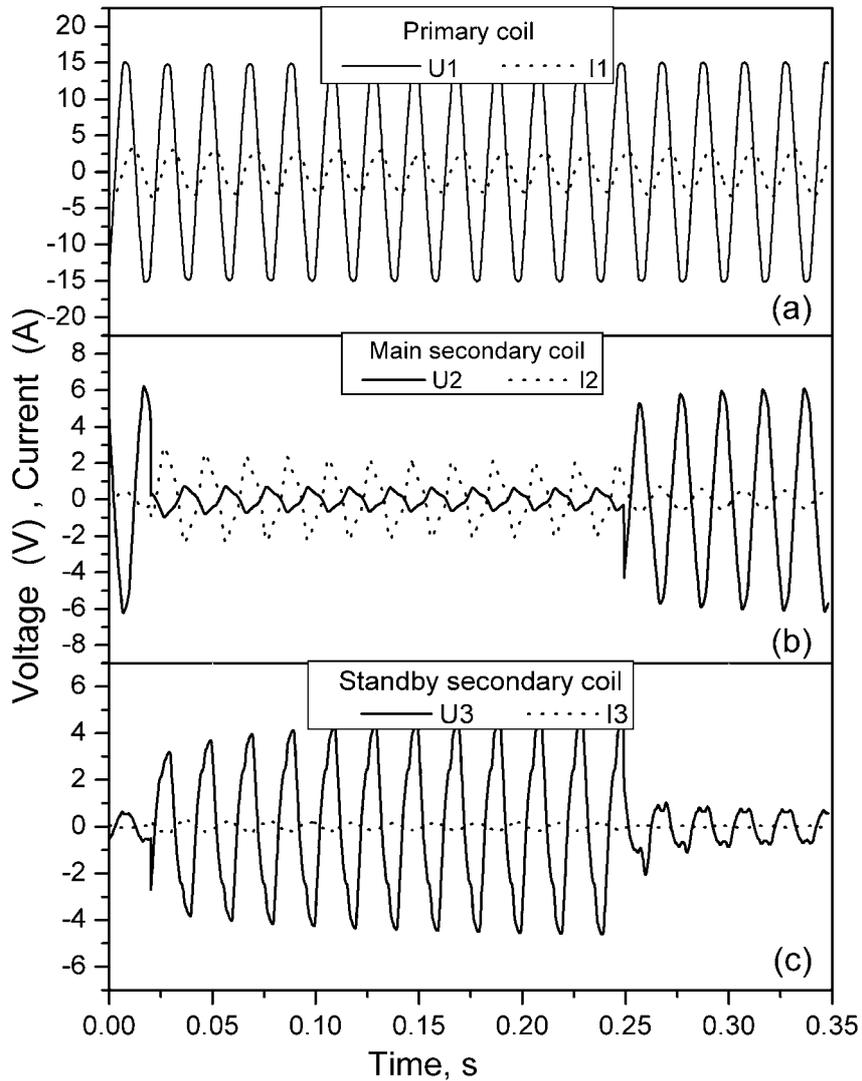

Fig. 4. Transient processes at a fault: (a) primary voltage $U_1$ and current $I_1$ (voltage and current are practically unchanged at a fault at 0.02 sec and at a fault clearing at 0.25 sec); (b) voltage $U_2$ at the terminals of the main secondary coil $W_2$-$W_2$' and current $I_2$ in the main line; (c) voltage $U_3$ and current $I_3$ in the standby line.